# The Effect of Students' Learning Orientations on Performance in Problem Solving Pedagogical Implementations

Charles A. Bertram[1,2] and Andrew J. Mason[1]

[1]University of Central Arkansas, Department of Physics and Astronomy,
201 S. Donaghey Avenue, Conway, AR 72035

[2]Cabot High School and Cabot Freshman Academy, Cabot, AR 72023

**Abstract:** Students' learning orientation, as applied towards learning problem solving, may be differentiated into learning a problem solving framework for its own sake, learning for the sake of doing well in the course, and non-goal-related considerations. In previous work, the role of learning orientation in student performance on a metacognitive problem solving exercise appeared to have some correlation with choice of major within life sciences, gains in conceptual understanding and attitudes towards problem solving skills. We examine a larger data set, taken from fourteen laboratory sections over six semesters of an introductory algebra-based physics course at the University of Central Arkansas, in which students worked on the problem solving exercise with their laboratory partners prior to a conceptually related laboratory exercise. We discuss the implications of analysis of the larger data set, in consideration of two arguably different IPLS-like populations.

## I. INTRODUCTION

Recently designed Introductory Physics in Life Science (IPLS) courses re-structure the curriculum to focus on topics that are more pertinent to life science majors [1-4]. Research-based pedagogies on certain aspects of physics learning, e.g. problem solving [2], are carefully considered for these purposes. Designing an IPLS course requires considerations for the background of the student population, e.g. the curricular differences in biology and health science major tracks, in the event that both types of majors must enroll in the same IPLS course [3-4]. Health science majors may have different expectations and different levels of preparation than do biology majors.

Previous research centered on an attempt to understand the effect of this background difference between majors, specifically in the context of potential effect of learning orientation [5-6]. Preliminary analyses were made of 1-3 semesters of an introductory algebra-based physics course with an IPLS-like population, whose student population consisted of an approximately equal number of biology majors and health science majors, who respectively hail from different colleges within the participating university. One identified potential difference is of the learning orientation of each student: either "learner-oriented," i.e. learning physical science content for its own sake, or "performance-oriented," i.e. learning the content for the sake of performing well in coursework [7]. The notion of learning orientation was applied to a more specific context of learning physics problem solving in a metacognitive exercise, reflecting upon one's strengths and struggles in the process of solving a context-rich problem in a collaborative environment with one's laboratory partners [8-9]. A potential relationship between learning orientation and pre-post survey performance, as well as overall course performance, was recognized [5]; however, learning orientation did not initially show much correlation to the group dynamics and use of epistemic games [10] during the solution attempt [6].

### A. Current Research Goals

We update the findings from Ref. [5] with additional data from a larger sample size, covering six different semesters' sections of the introductory algebra-based physics course in question. We explore whether the initial results of the learning orientation categories are robust, and whether there is any relationship with choice of major, both in terms of overall course grade and in terms of changes in attitudes towards learning physics. Discussion will center on the newer results' potential implications regarding IPLS-like student populations.

## II. PROCEDURE

### B. Lab Problem Solving Exercise

Six sections of a regional four-year state university's introductory algebra-based physics course, ranging from the Spring 2014 to Spring 2017 semester, were chosen for the study. Instruction was similar for all semesters, with two exceptions: 1) the textbook was changed after the Spring 2014 semester, and 2) the Spring 2017 semester's laboratory sections took place in a new teaching laboratory room. The course structure involved three 50-minute lecture sessions and one 3-

hour lab session per week. Each semester's course contained 48 or 72 students, divided into two or three 24-student laboratory sections.

During the first hour of the laboratory session each week, students conducted a lab group problem solving exercise, adapted from a previous study centered on metacognitive reflection on problem solving mistakes [11]. Lab groups typically consisted of two or three students each. For most semesters, these group sizes occurred naturally at each table; the exception was the Spring 2017 semester, for which only six tables were available in the classroom, requiring six groups of four students each. To preserve similar group dynamics of previous semesters, groups of four students during the Spring 2017 semester were instructed to work in pairs and then check their work with the other pairing at the table. Students worked on a context-rich problem [9] that was related to both the week's lecture material and to the experiment that would follow the problem solving exercise in the lab period. Students were permitted to use notebooks and textbooks as reference material for the solution attempt. During this time, the instructor provided feedback and hints to each group upon request.

At the end of the problem solving exercise, the instructor reviewed the problem's solution on the board. Typically, most lab groups could arrive at the solution, either on their own or with assistance, on most weeks. The review's aim was to consider aspects of the problem solution process that students might have missed. Students also used this review time to write reflections on how they did well on the problem, as well as on what aspects of the solution caused them to struggle. A self-diagnosis rubric [11] was provided for this purpose.

### C. Data Collection

During the last laboratory session of the semester, students submitted an in-class written response to the following survey question: "In what ways did you find this exercise useful towards learning the material in the course?" Written responses were transcribed into a spreadsheet, with identifying information about the students removed and replaced by ID numbers. This removed bias for two researchers to use inter-rater reliability in categorizing the responses into different learning orientations, typically finding initial agreement on classification for at least 80% of students per laboratory section, and quickly resolving any disagreements on the remaining students. If a student's response to the survey question was unclear, the researchers used the student's response to a follow-up question for clarification: "Do you have any suggestions to make this exercise more useful toward learning the material in the course?" The responses were categorized as follows [5-6]. "Framework"-oriented students typically discussed how the exercise explicitly helped them learn a problem-solving framework (e.g. help on a specific framework step, relating problem solving to the rest of the course material, understanding the framework as a whole). "Performance"-oriented students typically discussed how the exercise assisted them in performing well in other aspects of the course (e.g. exams, homework, or the ensuing lab activity). A third orientation, "Vague," included students whose responses did not focus on either learning goal (e.g. how they liked working in groups, discussing other aspects of the course, leaving no answer).

The CLASS survey [12] was administered in pre-post format on the first and last laboratory sessions of the semester, respectively. Students who failed the course, who were absent during either day that this data was collected, or who conspicuously did not take any of the surveys seriously, were omitted.

### III. RESULTS

### A. Different Majors vs. Different Orientations

Table 1 shows the distributions of students across both semesters into the "Framework," "Performance," and "Vague" orientations of students towards the problem solving exercise. Overall, 218 students, 176 of whom were either Biology or Health science majors, had complete sets of pre-post and survey data. Thirty-five students who had other majors in the same Natural Science college as Biology, i.e. "Other NS," are included for relative comparison. Seven students who were not science majors ("Non-Sci") took the course to satisfy a general education requirement.

TABLE 1. Students categorized into problem solving exercise orientations, per end-of-semester survey responses.

| Group (n)     | Framework | Performance | Vague |
|---------------|-----------|-------------|-------|
| Biology (91)  | 36        | 30          | 25    |
| Health (85)   | 32        | 28          | 25    |
| Other NS (35) | 5         | 18          | 12    |
| Non-Sci (7)   | 3         | 3           | 1     |
| All (218)     | 76        | 79          | 63    |

In this data set, Biology and Health science majors show a similar distribution among the three learning orientations, with slightly higher populations of Framework-oriented students and slightly lower populations of Vague-oriented students. In previous research [5], this similarity was not apparent due to a smaller sample size. The other Natural Science majors, on the other hand, were almost entirely Performance or Vague oriented, and in that way are dissimilar to Biology and Health science majors.

Table 2 shows the overall CLASS performance for the choice of majors as well as for learning orientations, using the sample sizes indicated in Table 1. Reported results are pretest scores, in terms of the percentage of expert-like responses, and normalized gains in this percentage. In all instances of reported p-values, a Levene test was first conducted to check for equality of variance, and one-way ANOVA between two groups was performed accordingly. In the vast majority of comparisons, variances were not statistically significant between groups ($p < 0.05$).

TABLE 2. Pretest scores, in terms of percentage of expert-like responses, and gains on the CLASS survey, by major and by learning orientation. See Table 1 for sample sizes. Statistically significant differences within rows or columns are respectively highlighted in bold; see text for details.

|  | Frame-work | Perfor-mance | Vague | All Orientations |
|---|---|---|---|---|
| Pretest – Percentage of Expert-like Choices | | | | |
| Biology | 59% | 58% | 63% | **60%** |
| Health | 54% | 50% | 48% | **51%** |
| OtherNS | 58% | 65% | 56% | **61%** |
| All | 57% | 57% | 56% | 57% |
| Overall Gains in Expert-like Attitudes | | | | |
| Biology | +0.08 | +0.01 | -0.02 | +0.03 |
| Health | ***+0.10*** | ***-0.06*** | ***-0.08*** | -0.00 |
| OtherNS | +0.19 | -0.02 | +0.09 | +0.04 |
| All | **+0.11** | **-0.03** | **-0.03** | +0.02 |

For all students in each learning orientation, there was no difference between groups on the pretest. However, there was a difference between learning orientations on overall attitudinal gains. Framework-oriented students experienced a non-trivial positive gain (SE = +/- 0.03 for all majors in each orientation). The other orientations each averaged a slight decline (albeit within SE of zero gain), which is similar to reported observations in large traditional-style courses [13].

For all students in each choice of major, there existed a strong difference in expert-like views of physics, with Health Science majors showing a significantly more novice-like attitudes towards physics than did Biology majors ($p < 0.01$) and other Natural Science majors ($p < 0.01$). In terms of gains, there were no significant differences between choice of major; the difference between life science majors in expert-like attitudes on the pretest still existed on the posttest.

Combining the two factors of learning orientation and choice of major in Table 2, more details arise as to differences between groups. Within each major, there is still no difference between learning orientations in pretest scores; the difference between majors on the pretest appears independent of learning orientation. The difference in gains between learning orientations appears to be prominent only within the Health Science majors (in italics; SE = +/-0.06 for all groups), in which the Framework-oriented Health Science majors had more expert-like gains than their Performance-oriented ($p < 0.05$) and Vague-oriented ($p < 0.01$) counterparts, both of which had slightly novice-like gains. This pattern does not emerge with Natural Science majors, biology or otherwise ($p > 0.10$, all comparisons between Framework-oriented groups and other groups within Biology and Other NS majors). Furthermore, it appears that Framework-oriented students will experience similar expert-like gains regardless of choice of major.

### B. Comparison to Course Performance

In consideration of whether learning orientation or choice of major has anything to do with overall course performance, Table 3 presents average GPAs for each set of categories, including p-values from t-test comparisons between groups. Students' course grades are interpreted as follows: A = 4.0, B = 3.0, C = 2.0, and D = 1.0. For example, the overall GPA for the entire student sample was 2.87, i.e. in the B-minus range (between 2.67 and 3.00).

TABLE 3. Average course grade for orientation groups and major groups, with standard error for each group, and p-values between groups.

| Semester | Average GPA | SE |
|---|---|---|
| All Students (218) | 2.87 | 0.06 |
| Framework (76) | 2.91 | 0.10 |
| Performance (79) | 2.97 | 0.09 |
| Vague (63) | 2.68 | 0.12 |
| Biology (91) | 3.18 | 0.08 |
| Health (85) | 2.58 | 0.10 |
| Other NS (35) | 2.80 | 0.14 |
| Frame vs. Perform | | $p = 0.63$ |
| Frame vs. Vague | | $p = 0.15$ |
| Perform vs. Vague | | *$p = 0.05$* |
| Biology vs. Health | | **$p < 0.00001$** |
| Biology vs. Other NS | | **$p < 0.02$** |
| Health vs. Other NS | | $p = 0.22$ |

By learning orientation, there is no significant difference between Framework-oriented groups and Performance-oriented groups. The Vague-oriented group is borderline significantly worse than the Framework-oriented group ($p = 0.15$) and the Performance-oriented group ($p = 0.05$), on the scale of a third of a letter grade. By choice of major, the difference between groups is more dramatic: the Biology majors, on average, perform much better than Health Science majors and other Natural Science majors do. In addition, all Natural Science majors (Biology and otherwise) average better than Health Science majors do ($p < 0.0001$). The overall results from Table 3 are generally upheld for within-groups comparisons based upon learning orientation. For each orientation, Biology majors perform better than do their Health Science counterparts ($p < 0.05$, all comparisons). In addition, no significant difference in course grade exists between learning orientations within each major ($p > 0.10$, all comparisons).

## IV. DISCUSSION

Framework-oriented students developed a more expert-like attitude over the course of the semester (as seen in Table 2), and appear to perform marginally better in the course (as seen in Table 3), than do Vague-oriented students. Performance-oriented students, meanwhile, do not become more expert-like than Vague-oriented students do, but still perform marginally better in the course. While biology and health science majors appear to have equivalent distributions of learning orientation (Table 1), the difference in attitudinal shifts by learning orientation seems to be most prominent within the Health Science majors, with Framework-oriented students becoming more expert-like and other students becoming more novice-like.

Natural Science majors, especially Biology majors, appear to have a significant advantage over Health Science majors, in terms of entering and finishing the course with more expert-like views of physics, and in terms of earning better overall grades in the course. A possible reason for this is that the Health Science majors reside in a different college than do the Natural Science majors. For example, Performance-oriented students' study habits may not be as beneficial for a course outside their home college as it is for a course within their home college. A potential way to address this apparent achievement gap between majors is to coordinate with members of the biology department as well as members of the health science college. This may require offering separate sections for each type of major that more specifically addresses each major's needs; however, difficulties may arise in terms of available faculty members to teach both sets of sections.

We note that learning orientations here are defined only with respect to a problem solving exercise within one aspect of the course. The problem solving items within the CLASS (to be addressed in a future study) are therefore likely the most valid measure related to learning orientations. Regardless, learning orientations show an effect on overall CLASS results, and even a marginal effect in overall course grade results. Literature in cognition suggests a stronger connection to mastery and performance achievement goals [14], including transfer of learning [15] to other aspects of the course, which may affect attitudes towards learning.

The study's results present potentially important considerations when designing IPLS courses that include both health science majors and life science majors [3-4], as well as a potential metric to interpret pre-post performance measures for these courses, e.g. whether measurements on an instrument like the CLASS are negative [13] or positive [16]. Future studies will more thoroughly define these considerations.

## ACKNOWLEDGEMENTS

We thank UCA's SPS program for contributions in collecting data and discussions about data analysis. We also thank T. Nokes-Malach for assistance with certain references. The University of Central Arkansas Sponsored Programs Office and Department of Physics and Astronomy provided funding for this project.[1] E.F. Redish et al., Am. J. Phys. **82**, 368 (2014).

[2] C.H. Crouch and K. Heller, Am. J. Phys. **82**, 378 (2014).

[3] S. Mochrie, Am. J. Phys. **84**, 542 (2016).

[4] E. Mylott, E. Kutschera, J. Dunlap, W. Christensen, and R. Widenhorn, J. Sci. Ed. Tech. **25**, 222 (2016).

[5] A. Mason, 2015 PERC Proceedings, edited by A. Churukian, D. Jones, and L. Ding, 215 (2015).

[6] A. Mason and C. Bertram, 2016 PERC Proceedings, edited by D. Jones, L. Ding, and A. Traxler, 224 (2016).

[7] Z. Hazari, G. Potvin, R.H. Tai, and J. Almarode, Phys. Rev. PER **6**, 010107 (2010).

[8] P. Heller, R. Keith, and S. Anderson, Am. J. Phys. **60**, 627 (1992).

[9] P. Heller and M. Hollabaugh, Am. J. Phys. **60**, 637 (1992).

[10] J. Tuminaro & E. Redish, Phys. Rev. PER **3**, 020101 (2007).

[11] E. Yerushalmi, E. Cohen, A. Mason, and C. Singh, Phys. Rev. PER **8**, 020109 and 020110 (2012).

[12] W. Adams et al., Phys. Rev. ST-PER **2**, 010101 (2006).

[13] S. Pollock, 2004 PERC Proceedings, edited by J. Marx, P. Heron, and S. Franklin, AIP **790**, 137 (2004).

[14] A. Elliott, Ed. Psychologist **34**, 169 (1999).

[15] D. Belenky and T. Nokes-Malach, J. of the Learning Sciences **21**, 399 (2012).

[16] M. Milner-Bolotin, T. Antimirova, A. Noack, and A. Petrov, Phys. Rev. PER **7**, 020107 (2011).